\documentclass{edp-jp4}
\usepackage{graphicx}
\begin{document}
\title{Synchronization and modularity in complex networks}
\author{Alex Arenas}
\address{Departament d'Enginyeria Inform{\`a}tica i Matem{\`a}tiques,
  Universitat Rovira i Virgili, 43007 Tarragona, Spain}
\author{Albert D\'{\i}az-Guilera}
\address{Departament de F{\'\i}sica Fonamental, Universitat de
  Barcelona, Mart{\'\i} i Franqu{\`e}s 1, 08028 Barcelona, Spain}

\maketitle
\begin{abstract} 
We investigate the connection between the dynamics of synchronization and the modularity on complex networks. Simulating the Kuramoto's model in complex networks we determine patterns of meta-stability and calculate the modularity of the partition these patterns provide. The results indicate that the more stable the patterns are, the larger tends to be the modularity of the partition defined by them. This correlation works pretty well in homogeneous networks (all nodes have similar connectivity) but fails when networks contain hubs, mainly because the modularity is never improved where isolated nodes appear, whereas in the synchronization process the characteristic of hubs is to have a large stability when forming its own community.
\end{abstract}

\section{Introduction}
\label{intro}

The theory of complex networks has reported major advances in the understanding of the networked substrate in which many natural, social and technological processes take place. Complex networks are representative of the intricate connections between elements in systems as diverse as the Internet and the WWW, metabolic networks, neural networks, food webs, communication networks, transport networks, and social networks \cite{watts,strogatz}. The availability of wide databases of entities (nodes) and relations (links) as well as the advances in computation have provided scientists with the necessary tools to unravel the statistical properties of complex networks \cite{rev1,rev2,rev3}. 

One of the subjects that has received more attention, in the recent years, is the detection and characterization of intermediate topological scales in their structure. In particular, the problem of detection of {\em community structure}, meaning the appearance of densely connected groups of vertices, with only sparser connections between groups, has been intensely attacked from the scientific community \cite{newmanepjb,jstat}. The most successful solutions, in terms of accuracy and computational cost required, are those based on the optimization of a magnitude called {\em modularity} proposed by Newman \cite{newgirvan} that allows the comparison of different partitionings of the network. The modularity of a given partition is, up to a multiplicative constant, the number of edges falling within groups minus the expected number in an equivalent network with edges placed at random. Given a network partitioned into communities, being $c_i$ the community to which node $i$ is assigned, the mathematical definition of modularity is expressed in terms of the adjacency matrix $A_{ij}$ and the total number of links $m=\frac{1}{2}\sum_i k_i$ where $k_i$ is the degree of node $i$ as

\begin{equation} 
Q=\frac{1}{2m}\sum_{ij} (A_{ij}-\frac{k_i k_j}{2m})\delta(c_i,c_j)
\label{Q}
\end{equation}

The search for the optimal (largest) modularity value is an NP-hard
problem \cite{brandes} that means that the space of possible partitions grows faster than any power of the system size. For this reason, a
heuristic search strategy is mandatory to restrict the search
space while preserving the optimization goal \cite{newfast,clauset,doye,rogernat,duch,newspect}.
Indeed, it is possible to relate the current optimization problem
for $Q$ with classical problems in statistical physics, e.g. the
spin glass problem of finding the ground state energy
\cite{potts}, where algorithms inspired in natural
optimization processes as simulated annealing
and genetic algorithms have been successfully
used. 

In a different scenario, physicists have largely
studied the dynamics of complex biological systems, and in
particular the paradigmatic analysis of large populations of
coupled oscillators \cite{winfree,strogatzsync,kurabook}. The
connection between the study of synchronization processes and
complex networks is interesting by itself. Indeed, the original
inspiration of Watts and Strogatz in the development of the
Small-World network structure\cite{watts} was to
understand the synchronization of cricket chirps. These
synchronization phenomena as many others e.g. asian fireflies
flashing at unison, pacemaker cells in the heart oscillating in
harmony, etc. have been mainly described under the mean field
hypothesis that assumes that all oscillators behave identically
and interact with the rest of the population. Recently, the
emergence of synchronization phenomena in these systems has been
shown to be closely related to the underlying topology of
interactions at mesoscopic scales\cite{atay}.

Here we analyze the effect of the community structure in the path towards synchronization. We study the dynamics towards synchronization in
several types of structured complex networks and find an evolving community structure based on the recruitment of groups of nodes towards complete synchronization.
We will also provide a connection between the emergence of synchronized groups and the way nodes
are grouped in some of the agglomerative methods of community detection based on the maximization
of the modularity, as defined in (\ref{Q}), \cite{jstat,newfast}.

The paper is
structured as follows: in section II we present the
synchronization model studied. In section III we describe a method
to construct synthetic networks with a well prescribed
hierarchical community structure. In section IV, we expose the
analysis of the route towards synchronization and their
relationship with the topological structure. Finally, we conclude with a discussion about the
communities revealed by synchronization processes in complex networks.

\section{The dynamical model}
The first successful attempt to understand synchronization
phenomena, from a physicist's perspective, was due to Kuramoto \cite{kurabook}, who analyzed a
model of phase oscillators coupled through the sine of their phase
differences. The model is rich enough to display a large variety
of synchronization patterns and sufficiently flexible to be
adapted to many different contexts \cite{conradrev}. The Kuramoto
model consists of a population of $N$ coupled phase oscillators
where the phase of the $i$-th unit, denoted by $\theta_i(t)$,
evolves in time according to the following dynamics
\begin{equation}
\frac{d\theta_i}{dt}=\omega_i + \sum_{j}
K_{ij}\sin(\theta_j-\theta_i) \hspace{0.5cm} i=1,...,N
 \label{ks}
\end{equation}
\noindent where $\omega_i$ stands for its natural frequency and
$K_{ij}$ describes the coupling between units. The original model
studied by Kuramoto assumed mean-field interactions $K_{ij}=K,
\forall i,j$. In absence of noise the long time properties of the
population are determined by analyzing the only two factors which
play a role in the dynamics: the strength of the coupling $K$
whose effect tends to synchronize the oscillators (same phase)
versus the width of the distribution of natural frequencies, the
source of disorder which drives them to stay away each other by
running at different velocities. For unimodal distributions, there
is a critical coupling $K_c$ above which synchronization emerges
spontaneously.

\subsection{Synchronization in complex networks}

Recently, due to the realization that many networks in nature have
complex topologies, synchronization studies have been extended to systems with heterogeneous 
connectivity patterns
\cite{barahona,motter1,yamir,hong,motter2,lee,munozprl,chavez}.
Usually, due to the complexity of the analysis in these cases some further
assumptions have been introduced. For instance, it has been a
normal practice to assume that the oscillators are identical.
In absence of disorder, i.e. if  $(\omega_i = \omega
~\forall i)$ there is only one attractor of the dynamics: the
fully synchronized regime where $\theta_i = \theta, ~\forall i$.
In this context the interest concerns not the final locked state
in itself but the route to the attractor. In particular, it has
been shown \cite{yamir2,kahng} that high densely interconnected
sets of oscillators (motifs) synchronize more easily that those
with sparse connections. This scenario suggests that for a complex
network with a non-trivial connectivity pattern, starting from
random initial conditions, those highly interconnected units
forming local clusters will synchronize first and then, in a
sequential process, larger and larger spatial structures also will
do it up to the final state where the whole population should have
the same phase \cite{gardenes}. We have shown \cite{arenas,arenas1} this process to occur at different time scales if a clear community structure exists.  Thus, the dynamical
route towards the global attractor reveals different
topological structures, indeed some of them very similar to those which represent
communities in partitions with high modularities. 

\subsection{Order parameter}

It is a normal practice to define, for the Kuramoto model, a
global `'order parameter" to characterize the level of entrainment
between oscillators. The normal choice is to use the following
complex-valued order-parameter
\begin{equation}
 r e^{i\psi}=\frac{1}{N}\sum_{j=1}^{N} e^{i\theta_{j}}.
\end{equation}
where $r(t)$, with $0\leq r(t)\leq 1$, measures the coherence of the
oscillator population, and $\psi(t)$ is the average phase.
However, this definition, although suitable for mean-field models
is not efficient to identify local dynamic effects. In particular
it does not give information about the route to the attractor
(fully synchronization) in terms of local clusters which is so
important to identify functional groups or communities. For this
reason, instead of considering a global observable, we define a
local order parameter measuring the average of the correlation
between pairs of oscillators
 \begin{equation}
\rho_{ij}(t)=<cos(\theta_i(t)-\theta_j(t))>
 \label{ro}
\end{equation}
\noindent where the brackets stand for the average over initial random phases.
The main advantage of this approach is that it allows to trace the time evolution of pairs of
oscillators and therefore to identify compact clusters of synchronized oscillators reminiscent
of the existence of communities.

In previous works \cite{arenas, arenas1} we have analyzed the dynamics towards synchronization
in different networks with community structure. From the average correlations between
pairs of oscillators ($\rho_{ij}$) we define a dynamical connectivity matrix. We consider that
two nodes are linked if their correlation is above some fixed threshold. In this way
we start with a system of disconnected nodes. As time goes on, nodes merge into groups 
until they form a single synchronized component, for a time long enough. 

For networks with a clear community structure, we have been able to identify the jumps of the 
number of connected components in time with the complete eigenvalue spectrum of
the Laplacian matrix, showing a striking similarity. Nevertheless, here we will focus in the relation 
between a magnitude that describes the quality of the community partitioning, the modularity
(\ref{Q}), and the relative stability of the dynamical structures that are formed in the merging
process described above.

\section{Networks}

In the present work we analyze the same type of networks than in \cite{arenas,arenas1}. Those are 
structured networks with a clear community structure. Some of them are homogeneous in degree and
are generalizations of the model networks proposed in \cite{girvannewman}  as a benchmark for
community detection algorithms.
Other networks have special nodes that act as hubs. For a detailed description and visualization
of the networks the reader is pointed to \cite{arenas1}.

The networks we analyze are:

\begin{itemize}

\item {\em Networks with 1 level of community with in-homogeneous distribution of community sizes}:
it is a kind of network that has been proposed as a better benchmark 
for community detection algorithms, since in real networks the community sizes are not
homogeneously distributed \cite{epjb}.

\item {\em Networks with two and three hierarchical levels of homogeneous communities}: 
This generalization was proposed \cite{arenas,arenas1} to show that the synchronization 
dynamics is able to
find communities at different levels. 

In general, to construct such a network, one takes a set of $N$ nodes
and divide it into $n_1$ groups of equal size; each of these
groups is then divided into $n_2$ groups and so on up to a number
of steps $k$ which defines the number of hierarchical levels of
the network. Then we add links to the networks in such a way that
at each node we assign at random a number of $z_{1}$ neighbours
within its group at the first level, $z_{2}$ neighbours within
the group at the second level and so on. There is a remaining
number of links that each node has to the rest of the network,
that we will call $z_{out}$. In this case it is easy to compute
the modularity of the partition \cite{girvannewman} at any level $l\le k$
\begin{equation}
Q_{n_1\cdot n_2 \cdot \ldots \cdot n_l}=\frac{z_l+\ldots  + z_k}{z_{out}+z_1+\ldots + z_k}-\frac{1}{n_1 \cdot n_2 \ldots \cdot n_l}
\end{equation}
and its numerical value tells us how good
as partition into a given community structure is. 

Here we will consider networks with 
two hierarchical levels, 256 nodes, and $n_1=n_2=4$;
this gives two possible partitions: one with 4 communities and the
other one with 16 communities. 
In the case of three levels we take  64 nodes and $n_1=n_2=n_3=2$, and hence there are three possible partitions, 2, 4, and 8 equal size communities. 

\item {\em Hierarchical networks with hubs}:
There is a set of self-similar deterministic networks
that has been used as an example of hierarchical scale-free
networks, proposed by Ravasz and Barabasi \cite{RB}. This type of
networks, apart from its hierarchical structure has some nodes
with a special role in terms of number of connexions (hubs) in
contrast to the networks discussed previously that are essentially
homogeneous in degree.

\end{itemize}

\section{Synchronization and community structure}

We have simulated the Kuramoto's model (\ref{ks}) with a constant natural frequency for the whole population on the above mentioned structured complex networks. According to the dynamics
described in Sect. 2 we analyze the evolution of the system in time by averaging over random initial 
phases distributed homogeneously in the range $[0,2\pi]$. Two oscillators are synchronized
when its correlation, given by eq. (\ref{ro}), is above some fixed threshold.

In Figs. 1-6 we plot the time evolution for a set of selected networks. In the bottom panels we represent the number of communities as a function of time (in a logarithmic scale). A community here is 
identified with a group of synchronized units. We use then this partition of the network into 
dynamical communities to compute the modularity in the top panel of the figures.

We observe that the partition corresponding to the most stable synchronization groups in time also have large values of modularity. This interesting effect is showing that the meta-stability of the synchronized groups is related to the specific topological structure where the dynamics takes place. Keeping in mind this idea, it is natural to propose a method for community detection based on the dynamics towards synchronization, however this should be carefully considered. The first problem we face is that the optimal partition into communities given by maximizing the modularity $Q$ does not corresponds exactly to the most stable conformation of groups of synchronization. We have used and heuristic algorithm to optimize $Q$ based on extremal optimization \cite{duch} obtaining larger values for the modularity than those presented by the synchronization communities for one of the networks, see Table 1. Still more striking that this difference is the observation of the communities that present larger stability in the Ravasz-Barabasi type networks. These hierarchical networks are characterized by the presence of hubs, the role of hubs in the synchronization process is very different that the role played by the rest of nodes. As shown in \cite{arenas} the equations involving hubs in the synchronization process are topological averages of the phases of the nodes they are connected with. In terms of meta-stable patterns of synchronization, hubs persist during long times as isolated communities.  However, this fact could never be detected via optimization of the modularity, because modularity of any partition with isolated nodes can never be optimal. This last fact is proved analytically from the definition of modularity.

\section{Conclusions}

We have shown that meta-stable patterns of synchronization in the path towards complete synchronization are closely related to the partitions obtained optimizing modularity on complex networks. However, the correspondence between both descriptions is not exact. This is pointing out that some definition of communities at different scales from topological analysis, including the possibility of having nodes forming its own community will be more representative of the topological role of the structure in the dynamics taking place on it.  

\section{Acknowledgments}

This work has been supported by DGES of the Spanish Government Grants No. BFM-2003-08258 and FIS-2006-13321.

\begin{table}
\begin{tabular}{|l||r|r||r|r|}
\hline
&\multicolumn{2}{c||}{Extremal}&\multicolumn{2}{c|}{Synchronization}\\
\hline
& $Q_{max}$ & comms & $Q_{max}$ & comms \\
\hline
13\_4 & 0.696  & 4 & 0.696 & 4 \\
\hline
15\_2 & 0.772  & 16 & 0.772 & 16 \\
\hline
3n64 & 0.714  & 6 & 0.714 & 6 \\
\hline
inhomo & 0.609  & 6 & 0.609 & 5 \\
\hline
RB 25 & 0.551  & 5 & 0.551 & 5 \\
\hline
RB 125 & 0.642  & 11 & 0.626 & 9 \\
\hline
\end{tabular}
\caption{Maximum modularity and number of communities in the corresponding partition
obtained with our dynamical method and with the Extremal Optimization method \cite{duch}.
Labels for the networks correspond to those in Figs. 1-6, in this order.}
\end{table}

\begin{figure}
\resizebox{0.70\columnwidth}{!}{%
\includegraphics[angle=270]{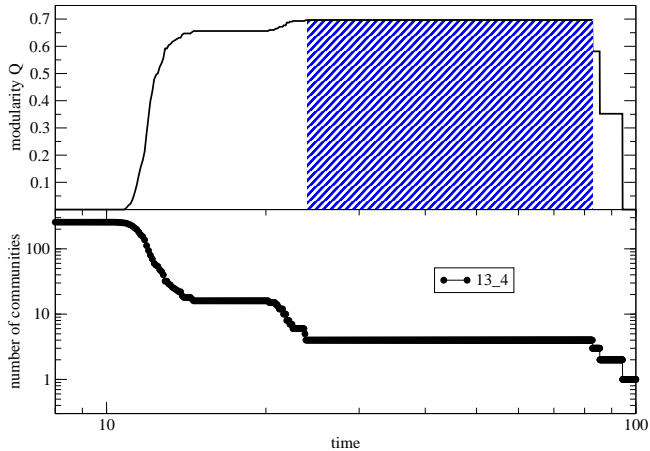}
}
\caption{Merging of groups of oscillators in time for a network of 256 nodes grouped in 2 hierarchical levels of communities, with $z_1$=13, $z_2=4$, and $z_{out}$=1. Bottom: Evolution of the number
of communities, identified as groups of synchronized oscillators. Top: modularity computed according 
to the partition given by the synchronized groups. The shadow area corresponds to
the largest modularity. Time is in a logarithmic scale and units are arbitrary.}
\label{13_4_fig}       
\end{figure}

\begin{figure}
\resizebox{0.70\columnwidth}{!}{%
\includegraphics[angle=270]{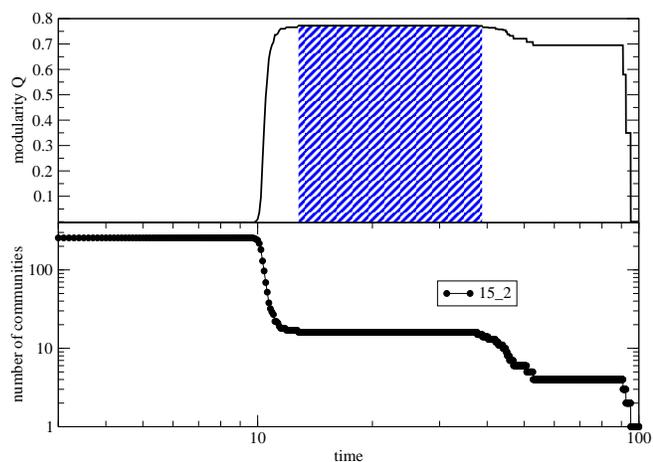}
}
\caption{Same as Fig. 1 for a network with $z_1$=15, $z_2=2$, and $z_{out}$=1.}
\label{15_2_fig}       
\end{figure}

\begin{figure}
\resizebox{0.70\columnwidth}{!}{%
\includegraphics[angle=270]{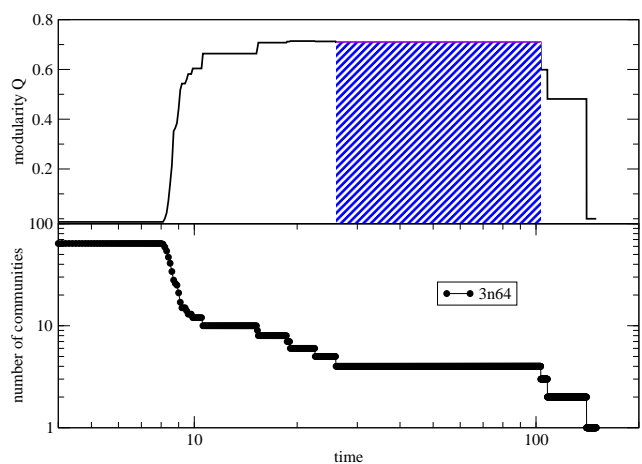}
}
\caption{Same as Fig. 1 for a network of 3 hierarchical levels of two branches each, with 64 nodes.}
\label{3n_fig}       
\end{figure}

\begin{figure}
\resizebox{0.70\columnwidth}{!}{%
\includegraphics[angle=270]{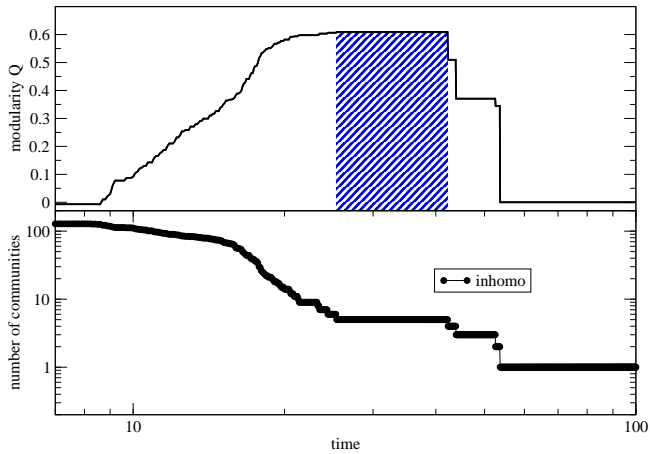}
}
\caption{Same as Fig. 1 for a network with 4 communities of 16 nodes each plus an additional one of
64 nodes.}
\label{inhomo_fig}       
\end{figure}

\begin{figure}
\resizebox{0.70\columnwidth}{!}{%
\includegraphics[angle=270]{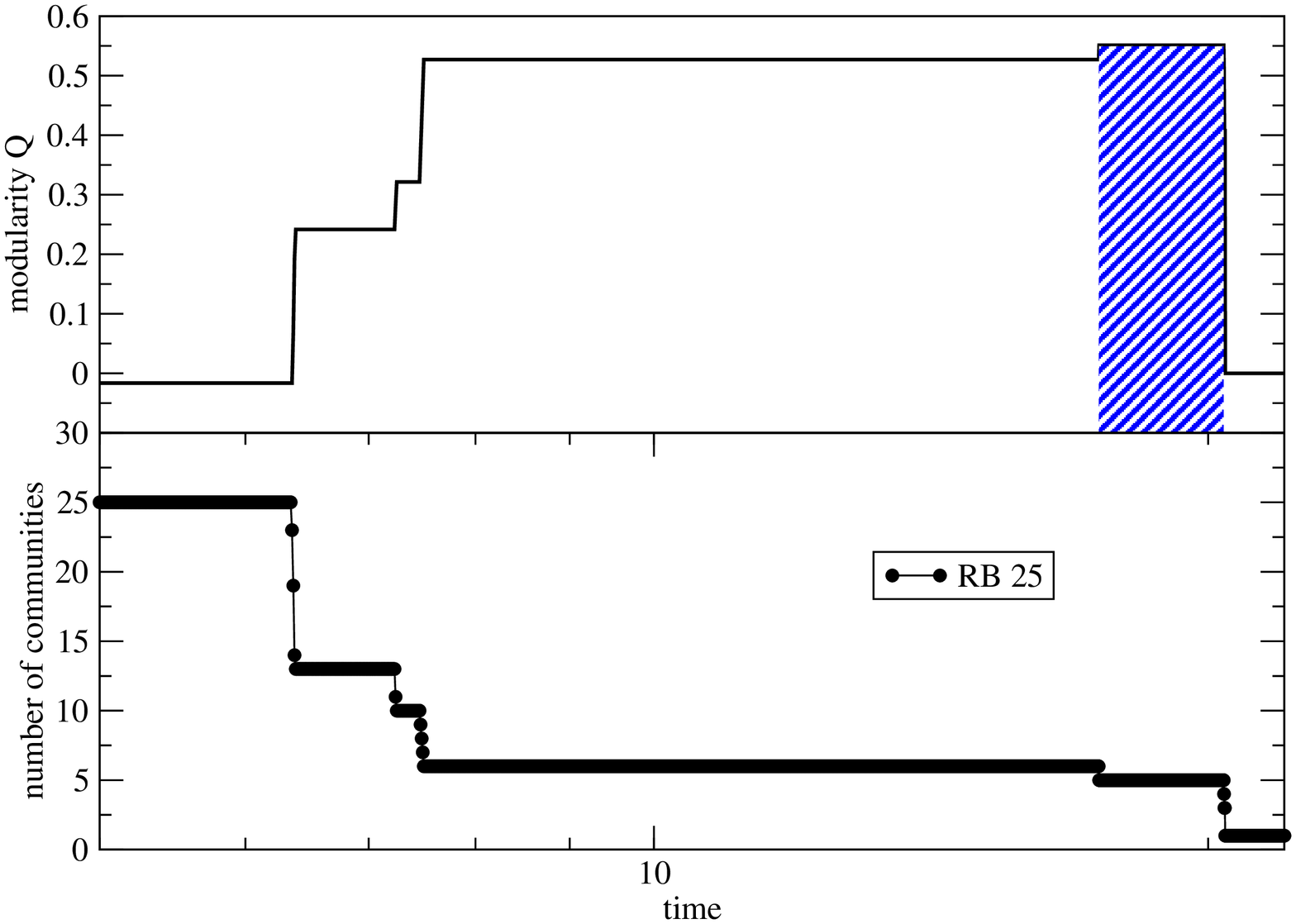}
}
\caption{Same as Fig. 1 for a hierarchical Ravasz-Barabasi network of 25 nodes.\cite{RB}}.
\label{RB_fig}       
\end{figure}

\begin{figure}
\resizebox{0.70\columnwidth}{!}{%
\includegraphics[angle=270]{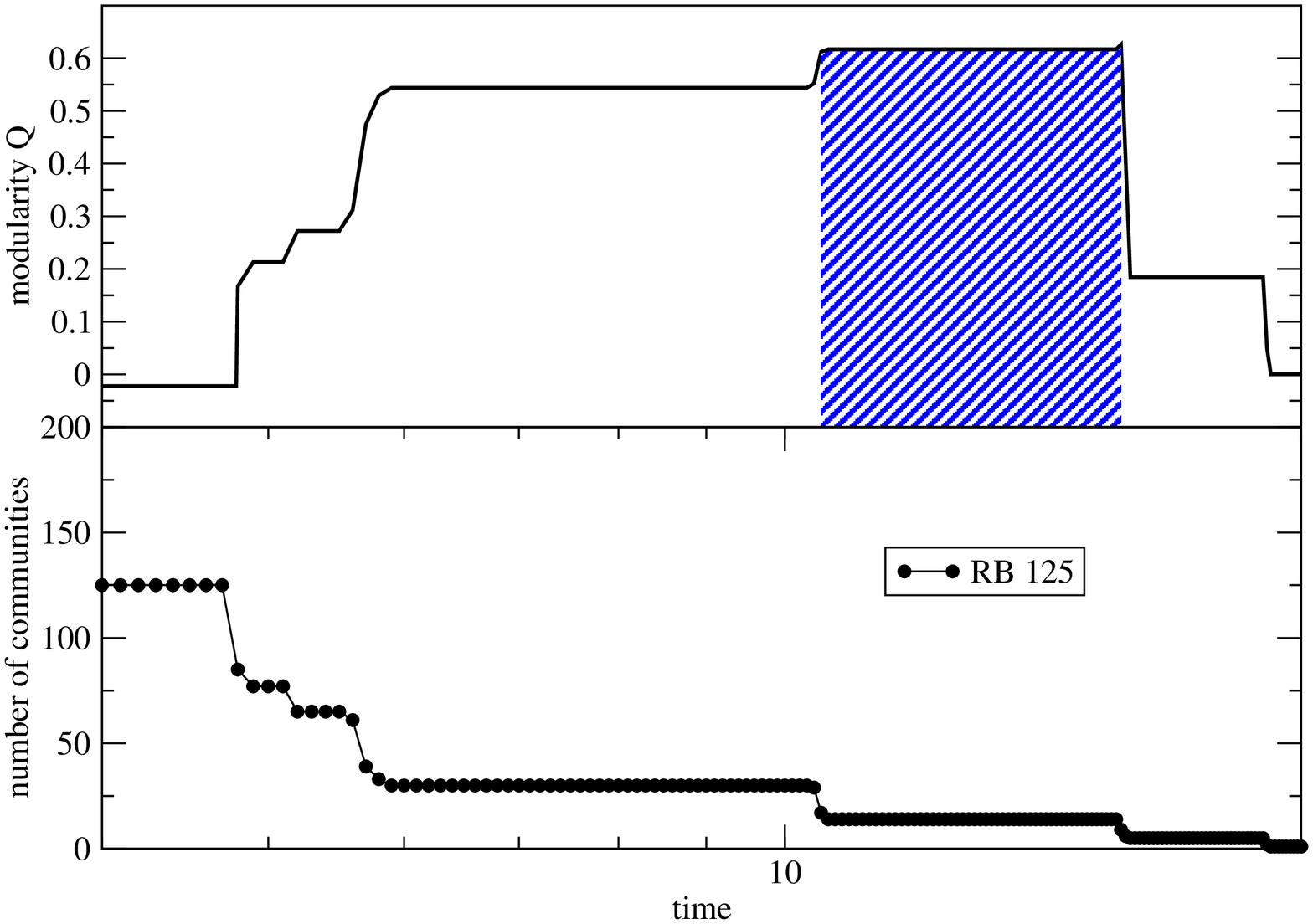}
}
\caption{Same as Fig. 1 for a hierarchical Ravasz-Barabasi network of 125 nodes.\cite{RB}}.
\label{RB3_fig}       
\end{figure}

\end{document}